\documentstyle[aps,12pt]{revtex}

\begin{document}
\draft
\author{Jerzy Matyjasek$^{1}$ and O. B. Zaslavskii$^{2}$}
\address{$^{1}$Institute of Physics, Maria Curie-Sk\l odowska University,\\
pl. Marii Curie-Sk\l odowskiej 1, 20-031 Lublin, Poland\\
E-mail address: matyjase@tytan.umcs.lublin.pl\\
$^{2}$Department of Mechanics and Mathematics,\\
Kharkov V.N. Karazin's National University,\\
Svoboda Sq.4, Kharkov 61077, Ukraine\\
E-mail address: ozaslav@kharkov.ua}
\title{Extremal limit for charged and rotating 2+1--dimensional black holes and
Bertotti-Robinson geometry }
\maketitle

\begin{abstract}
We consider 2+1--dimensional analogues of the Bertotti-Robinson (BR)
spacetimes in the sense that the coefficient at the angular part is a
constant. We show that such BR-like solutions are either pure static or
uncharged rotating. We trace the origin of the inconsistency between a
charge and rotation, considering the BR-like spacetime as a result of the
limiting transition of a non-extremal black hole to the extremal state. We
also find that the quasilocal energy and angular momentum of such BR-like
spacetimes calculated within the boundary $l=const$ ($l$ is the proper
distance) are constants independent of the position of the boundary.
\end{abstract}

\pacs{PACS numbers: 04.60.Kz, 04.20.Jb}

\baselineskip=19pt


\section{Introduction}

In recent years, interest to $AdS_{2}\times S_{2}$ geometries increased in
the context of string theory and AdS/CFT correspondence~\cite{3,w}. Apart
from this, the geometries of this type appear naturally in black hole
physics. If one considers a charged black hole and makes limiting transition 
$T_{H}\rightarrow 0$ ($T_{H}$ is the Hawking temperature) from the
non-extremal black hole geometry to the extremal state, such that the
canonical gravitating thermal ensemble remains well-defined~\cite{prl,extr},
the Reissner-Nordstr\"{o}m (RN) metric turns into the Bertotti-Robinson (BR)
spacetime \cite{r}, \cite{b} with the black hole horizon turning into the
acceleration one. The similar geometries are also relevant for non-linear
electrodynamics~\cite{jext}, string dust sources~\cite{dad},
higher-dimensional spacetimes~ \cite{vanzo,Oscar}, and quadratic gravity 
\cite{jt} The thermodynamic properties of acceleration horizons are
considered from a general viewpoint in~\cite{ac}. Moreover, it should be
emphasised that the limiting procedure is defined not only for
spherically-symmetrical spacetimes but also for generic static black hole
configurations~\cite{hm}, and, in particular, can be applied to different
versions of C-metric~\cite{jose}. In a similar way, the rotating analogs of
BR spacetimes are obtained from the Kerr solutions in Ref.~\cite{hm} for the
non-extremal horizons and in \cite{bard} for the extremal ones. Such
solution appear naturally in the context of the dilaton-axion gravity \cite
{gc}.

For 2+1 rotating uncharged black holes the limiting procedure under
consideration has been carried out in \cite{extr}. The resulting solution
coincides with that found in \cite{c4}, \cite{ch}. For the charged
unrotating 2+1 black holes the general procedure of \cite{extr} applies as
well and gives solutions found in \cite{c2} and discussed also in \cite{c5}.

The aim of this paper is to elucidate, whether such BR-like 2+1-dimensional
solutions exist when both rotation and charge are present. (By BR-like we
mean the $AdS_{2}\times S_{1}$ structure, the coefficient at the angular
part being a constant.) In this case the original black hole solutions, to
which the limiting procedure should apply, become rather complicated \cite
{c2}, \cite{c3}, \cite{q}. Meanwhile, the advantage of the general limiting
procedure elaborated in \cite{hm} consists just in the fact that it enables
to guess the general form of the metric. Therefore, instead of analysing the
original solutions with the subsequent applying the limiting transition, we
start from the Einstein equations directly in which we substitute the
anticipated form of the metric. As we will see below, such an approach
enables us to find at once a quite unexpected result: when the charge and
rotation are both present, BR-like geometries are impossible.

\section{Field equations}

Consider stationary 2+1--dimensional geometry described by a line element of
the form 
\begin{equation}
ds^{2}=-N^{2}(\rho )f^{2}(\rho )dt^{2}+f(\rho )^{-2}d\rho ^{2}+r^{2}(\rho
)\left( d\phi +N^{\phi }(\rho )dt\right) ^{2}\text{. }  \label{m}
\end{equation}
Of matter source we assume that it is purely electromagnetic, with the
stress-energy tensor given by 
\begin{equation}
8\pi T_{\mu }^{\nu }\equiv \theta _{\mu }^{\nu }=2F^{\mu \alpha }F_{\nu
\alpha }-\frac{1}{2}\delta _{\mu }^{\nu }F_{\alpha \beta }F^{\alpha \beta }%
\text{,}  \label{set}
\end{equation}
where $F_{\alpha \beta }$ is the electromagnetic tensor. We adopt the
notations $t=x^{0}$, $\rho =x^{1}$, $\phi =x^{2}$.

For the electromagnetic tensor compatible with the assumed symmetries, it
follows from the Maxwell equations that $F^{01}=-\frac{Q}{Nr}$, $F^{21}=%
\frac{P}{Nr}$,where the constants $Q$ and $P$ have the meaning of an
electric and magnetic charge, respectively.

The metric (\ref{m}) preserves its form under the coordinate transformation $%
\phi =\phi ^{\prime }+\Omega t$, where $\Omega $ is a constant. In doing so, 
$\left( N^{\phi }\right) ^{\prime }=N^{\phi }+\Omega $ and $\left(
F^{21}\right) ^{\prime }=F^{21}-F^{01}\Omega $. As both $F^{21}$ and $F^{01}$
have the same coordinate dependence, one can always achieve $F^{12}=0$,
choosing $\Omega =-\frac{P}{Q}$. In what follows we assume that this
condition is satisfied. Then, the only nonvanishing components of $\theta
_{\nu }^{\mu }$ are simply $\theta _{0}^{0}=Q^{2}\left[ \frac{\left( N^{\phi
}\right) ^{2}}{f^{2}N^{2}}-\frac{1}{r^{2}}\right] =\theta _{1}^{1}=-\theta
_{\phi }^{\phi }$, and $\theta _{\phi }^{0}=\frac{2Q^{2}N^{\phi }}{N^{2}f^{2}%
}$. It is worth noting that the regularity of the stress-energy tensor on
the event horizon requires $N^{\phi }\sim f^{2}\rightarrow 0$. Thus, our
frame turns out to be corotating with the horizon automatically.

For the line element (\ref{m}) it is convenient to use the following
combinations of Einstein equations (where the cosmological constant $\Lambda
=-\left| \Lambda \right| <0$): 
\begin{equation}
G_{t}^{t}-N^{\phi }G_{\phi }^{t}=\theta _{t}^{t}-N^{\phi }\theta _{\phi
}^{t}-\Lambda \text{,}  \label{g15}
\end{equation}
\begin{equation}
NG_{\phi }^{t}=N\theta _{\phi }^{t}\text{,}  \label{g16}
\end{equation}
\begin{equation}
G_{\phi }^{t}N^{\phi }-G_{t}^{t}+G_{\rho }^{\rho }=\theta _{\phi
}^{t}N^{\phi }-\theta _{t}^{t}+\theta _{\rho }^{\rho }\text{,}  \label{g17}
\end{equation}
\begin{equation}
G_{t}^{t}+G_{\phi }^{\phi }=\theta _{t}^{t}+\theta _{\phi }^{\phi }-2\Lambda 
\text{.}  \label{ad}
\end{equation}

Henceforth, we shall use the proper length $l$ as a radial coordinate,
i. e., we will work in the gauge 
\begin{equation}
ds^{2}=-M^{2}dt^{2}+dl^{2}+r^{2}(l)\left( d\phi +N^{\phi }(l)dt\right) ^{2}%
\text{,}  \label{new}
\end{equation}
where $M=Nf$ . Then in this system we can rewrite equations (\ref{g15}--\ref
{ad}) in the form:

\begin{equation}
2\frac{d^{2}r}{dl^{2}}+2\Lambda r+\frac{1}{2}\frac{r^{3}}{M^{2}}\left( \frac{%
dN^{\phi }}{dl}\right) ^{2}+2\frac{Q^{2}}{r}+\frac{2\left( N^{\phi }Q\right)
^{2}r}{M^{2}}=0\text{,}  \label{015}
\end{equation}
\begin{equation}
\frac{d}{dl}\left( \frac{r^{3}}{M}\frac{dN^{\phi }}{dl}\right) =2\frac{%
Q^{2}N^{\phi }r}{M}\text{,}  \label{016}
\end{equation}
\begin{equation}
\frac{1}{M}\frac{dM}{dl}\frac{dr}{dl}-\frac{d^{2}r}{dl^{2}}=2\frac{r\left(
QN^{\phi }\right) ^{2}}{M^{2}},  \label{017}
\end{equation}
\begin{equation}
\frac{1}{M}\frac{d^{2}M}{dl^{2}}+\frac{1}{r}\frac{d^{2}r}{dl^{2}}-\frac{1}{2}%
\frac{r^{2}}{M^{2}}\left( \frac{dN^{\phi }}{dl}\right) ^{2}+2\Lambda =0.
\label{018}
\end{equation}

\section{{{{{{{{{\protect\large {\lowercase{$r(\l)$}=\lowercase{$const$} 
\lowercase{$=r_{0}$}}}}}}}}}}}

Eqs. (\ref{015}) - (\ref{017}) describe two qualitatively different
situations. If $r(l)$ is not constant identically, Eqs.~(\ref{015}) - (\ref
{017}) comprise the full set of three independent equations for three
unknown functions $r(l)$, $N(l)$ and $N^{\phi }(l)$. Remaining equations, as
for instance, $(_{\phi }^{\phi })$ equation, can be easily obtained from
them with the help of Bianchi identities. However, if $r(l)=const\equiv r_{0}
$, only two equations of (\ref{015}) - (\ref{017}) are independent. Indeed,
it could be easily demonstrated that the left hand side of Eq.~(\ref{017})
identically vanishes, and, consequently, 
\begin{equation}
QN^{\phi }=0\text{.}  \label{qn}
\end{equation}
As the equation for determining $M$ is now missing, Eqs. (\ref{015}) - (\ref
{017}) are insufficient for the case $r=r_{0\text{ }}$and should be
supplemented by the additional equation, as for example Eq. (\ref{ad}) or,
equivalently, (\ref{018}). It is worth stressing that, being the consequence
of Eqs. (\ref{015}) - (\ref{017}) in the case $r(l)\,\neq const$, now it
represents a new independent equation, which in the present context gives 
\begin{equation}
\frac{1}{M}\frac{d^{2}M}{dl^{2}}=\alpha ^{2}\text{, \ \ \ \ }\alpha \equiv 
\frac{r_{0}}{M}\frac{dN^{\phi }}{dl}=\frac{2}{L}\text{. }  \label{m''}
\end{equation}

It follows from Eq. (\ref{qn}) that the general case is splitted to two
subcases. Although, as is mentioned in Introduction, the solutions for each
of them are known, for completeness we rederive them below in a rather
straightforward manner.

1) $Q=0$, $N^{\phi }\neq 0$.

Then

\begin{equation}
M=\frac{L}{2}m(\alpha l)\text{, \ \ \ }N^{\phi }=\frac{L}{2r_{0}}n(\alpha l)%
\text{, }  \label{M}
\end{equation}
where we choose the normalization of time (i.e. the coefficient at $M$) in
accordance with the limiting transition for 2+1 black holes to the extremal
state (see below). There are three physically different solutions ($x\equiv
\alpha l$): (1a) $m(x)=\sinh x$, $n(x)=2\sinh ^{2}\frac{x}{2}$. (1b) $m=\exp
(x)$, $n=\exp (x)$, (1c) $m=\cosh x$, $n=\sinh x$. The cases (1a) - (1c)
correspond to the solutions found in \cite{c4}, \cite{ch}. The case (1a) can
be also obtained by taking the extremal limit of the non-extremal 2+1 black
hole \cite{extr}.

2) $Q\neq 0$, $N^{\phi }=0$.

It follows from Eq. (\ref{015}) 
\begin{equation}
r_{0}^{2}=Q^{2}L^{2}\text{.}
\end{equation}
Eq. (\ref{m''}) with $\alpha =0$ gives us again three possibilities (2a) $M=%
{\displaystyle{\sinh al \over a}}%
$, (2b) $M=\cosh al$, (2c) $M=\exp (al)$,where $a^{2}=\frac{2}{L^{2}}$. In
the limit $l\rightarrow \infty $ ($\Lambda \rightarrow 0$), $Q\rightarrow 0$%
, $r_{0}=const$ we obtain the flat spacetime. The solutions (2a) - (2c)
correspond to those found in \cite{c2}.

All solutions (1a--1c) and (2a--2c) share a rather peculiar property which
was not noticed before. One can calculate the quasilocal energy density $%
\varepsilon =\frac{k}{8\pi }+\varepsilon _{0}$, where $\varepsilon _{0}$
comes from the contribution of the reference background, $k$ is the
extrinsic curvature of the boundary embedded into the three-dimensional
spacelike surface \cite{by}. Choosing the foliation $t=const$ and the
boundary at fixed $l$, we find that the only non-vanishing component of the
unit normal to the boundary is $n^{1}=1$ and $k=0.$ Thus, the energy density 
$\varepsilon =\varepsilon _{0}(r_{0})=const$ since $r_{0}=const$. This is
the typical feature of acceleration horizons \cite{ac}.

In a similar way one can calculate the quasilocal angular momentum \cite{jm} 
\begin{equation}
J_{B}=\frac{1}{\pi }\int d\sigma K_{ij}\xi ^{i}n^{j}\text{,}
\end{equation}
$d\sigma $ is the proper element of the boundary, $\xi ^{i}$ is the axial
Killing vector, the extrinsic curvature tensor 
\begin{equation}
K_{ij}=\frac{1}{2N}(N_{i\mid j}+N_{j\mid i})\text{,}
\end{equation}
$N_{i\mid j}$ is the covariant derivative of the shift vector with respect
to the slice $t=const$. Then, direct calculation gives us for the boundary
with an arbitrary fixed $l$ that 
\begin{equation}
J_{B}=2r_{0}^{2}/L.  \label{j1}
\end{equation}
Thus, both the energy and angular momentum turned out to be constant.

\section{Limiting transition}

As we saw in the preceding sections, BR-like configurations with
simultaneously nonzero $N^{\phi }$ and $Q$ are impossible. To understand
better this fact, we\ shall employ the approach based on the limiting
transition \cite{extr} relying only on the structure of field equations. For
the metric (\ref{m}) one can always achieve $N^{\phi }=0$ on the horizon
passing to the frame corotating with a black hole. Using in Eq. (\ref{m})
the gauge $r=\rho $ and assuming for simplicity $N=1$, we can write the
asymptotic expansion of metric potentials near the extremal state in the
form 
\begin{equation}
f^{2}=4\pi T_{H}(\rho -\rho _{+})+b(\rho -\rho _{+})^{2}\text{, \ \ \ \ \ }%
N^{\phi }=c(\rho -\rho _{+})
\end{equation}
and 
\begin{equation}
\rho -\rho _{+}=4\pi T_{H}b^{-1}\sinh ^{2}\frac{x}{2}\text{, \ \ \ \ \ \ \ }%
t=\frac{\tilde{t}}{2\pi T_{H}}\text{.}
\end{equation}
In the limit $T_{H}\rightarrow 0$ we obtain 
\begin{equation}
ds^{2}=b^{-1}\left( -\sinh ^{2}xd\tilde{t}^{2}+dx^{2}\right) +\rho
_{+}^{2}(d\phi +d\sinh ^{2}\frac{x}{2}d\tilde{t})^{2}\text{,}  \label{2+1bh}
\end{equation}
where $d=2cb^{-1}$.

In particular, for 2+1 black holes \cite{2+1}, \cite{2+1d} 
\begin{equation}
f^{2}=\frac{\rho ^{2}}{L^{2}}-M+\frac{J^{2}}{4\rho ^{2}}\text{, }b^{-1}=%
\frac{L^{2}}{4}
\end{equation}
and 
\begin{equation}
N^{\phi }=\frac{J}{2}\left( \frac{1}{\rho _{+}^{2}}-\frac{1}{\rho ^{2}}%
\right) \text{,}  \label{j}
\end{equation}
where the constant $J$ represents the angular momentum. We see that $c=\frac{%
J}{\rho _{+}}$. In the extremal limit 
\begin{equation}
J=\frac{2\rho _{+}^{2}}{L}\text{, }c=\frac{2\rho _{+}}{L\text{ }}\text{, }d=%
\frac{L}{\rho _{+}}\text{.}  \label{ang}
\end{equation}
For uncharged black holes, the metric (\ref{2+1bh}) corresponds to the case
(1a) and agrees with Eq. (22) of Ref. \cite{extr} (with typographical errors
corrected). However, if $Q\neq 0$, the quantity $N\neq 1$ \cite{q}. It
follows from eqs. (17), (19) of \cite{q} or directly from eq. (\ref{g17})
with $r(\rho )=\rho $ that 
\begin{equation}
\frac{d}{d\rho }N=\frac{\left( QN^{\phi }\right) ^{2}\rho }{2Nf^{4}}\text{.}
\label{17}
\end{equation}
Therefore, we see from (\ref{17}) that near the horizon $N^{\phi }\sim
f^{2}\sim T_{H}(\rho -\rho _{+})$. Thus, the coefficient $c$ is proportional
to $T_{H}$ and vanishes as $T_{H}\rightarrow 0$. As in the process of the
limiting transition the radial coordinate of all points of the manifold
approaches the horizon value $\rho =\rho _{+}$, it turns out that the
coefficient $d\rightarrow 0$, so the system becomes static. Thus, the
interpretation of the metrics under discussion as extremal limits of
corresponding non-extremal configuration explains why the limiting
configuration cannot be simultaneously rotating and charged.

It is worth noting that for 2+1 black holes the angular momentum $%
J_{B}=const=J$ \cite{en2+1}, \cite{bcm}. Therefore, the fact that $J=const$
also for the limiting form of the metric (established in the previous
Section) is not surprising. In doing so, eq.(\ref{ang}) with $\rho
_{+}=r_{0} $ agrees with (\ref{j1}). However, for 2+1 black holes the energy
is not a constant \cite{en2+1}, \cite{bcm}. Nevertheless, after the limiting
transition under discussion it becomes constant. Indeed, calculating the
quasilocal energy for the boundary $\rho =r_{0}(l_{0})$, one obtains $%
\varepsilon =-\frac{1}{4\pi r_{0}}f(r_{0})+\varepsilon _{0}(r_{0})$, where $%
\varepsilon _{0}(r_{0})$ is the subtraction term. After the limiting
transition $r_{0}=\rho _{+}=const$, so that $f(r_{0})=0$ and we obtain that $%
\varepsilon =\varepsilon _{0}$ does not depend on the distance from the
horizon. Thus, for the solution with an acceleration non-extremal horizon
(1a) and (2a) the limiting procedure under discussion explains also the
property of the energy indicated at the end of Section IV.

\section{Summary}

In this note we established that 2+1 BR-like geometries ($r(\rho
)=r_{0}=const$)\ can exist only for the rotating or charged case separately
but are forbidden when rotation and charge are present simultaneously.
(Throughout the paper we assumed that our frame is corotating with the
horizon, so one can make coordinate transformations to the new system
rotating with a constant angular velocity but such a transition is a trivial
and gives no new solutions.) This fact is in a sharp contrast with the 3+1
case when generalizations of BR-like solutions do exist and can be obtained
form the extremal Kerr-Newman geometries by the suitable limiting transition 
\cite{hm}, \cite{bard}. As the corresponding metrics with extremal horizons
describe the near-horizon region of the extremal black holes, the property
under discussion may have important consequences for the late stage of 2+1
collapse to the extremal state, if rotation is supplemented by the presence
of a charge, however small it be.

It is also worth noting that the BR-like configurations under discussion
turned out to be ``poor'' in the dynamic sense since both the angular
momentum (for $Q=0$, $N^{\phi }\neq 0$) and the energy calculated within the
boundary $l=const$ $=l_{B}$ do not depend on $l_{B}$. This is in agreement
with the symmetry of the system according to which the geometrical
properties of sections with different $l_{B}$ including the horizon are
equivalent.

\section{Acknowledgment}

The authors thank G\'{e}rard Cl\'{e}ment for useful correspondence.





%
%

%
%

\end{document}